\documentclass[a4paper,fleqn]{cas-dc}

\usepackage[numbers]{natbib}
\usepackage{enumitem}
\usepackage[ruled,vlined]{algorithm2e}
\usepackage{diagbox}
\usepackage{xcolor} 
\def\tsc#1{\csdef{#1}{\textsc{\lowercase{#1}}\xspace}}
\tsc{WGM}
\tsc{QE}
\tsc{EP}
\tsc{PMS}
\tsc{BEC}
\tsc{DE}

\begin{document}
\let\WriteBookmarks\relax
\def\floatpagepagefraction{1}
\def\textpagefraction{.001}

\shorttitle{FedDiSC: A Computation-efficient Federated Learning Framework for Power Systems Disturbance and Cyber Attack Discrimination}

\shortauthors{M. A. Husnoo et~al.}

\title [mode = title]{FedDiSC: A Computation-efficient Federated Learning Framework for Power Systems Disturbance and Cyber Attack Discrimination}                      
\tnotemark[1]
\tnotetext[1]{This work is the result of the research project funded by the Faculty of Science, Engineering and Built Environment (SEBE), Deakin University under the scheme 'Mini ARC Analogue Program (MAAP)'.}

\author[1]{Muhammad Akbar Husnoo}[orcid=0000-0001-7908-8807]
\cormark[1]
\fnmark[2]
\ead{mahusnoo@deakin.edu.au}

\address[1]{Centre for Cyber Security Research and Innovation (CSRI), School of Information Technology, Deakin University, Geelong, VIC 3216, Australia}

\author[1]{Adnan Anwar}[orcid=0000-0003-3916-1381]
\ead{adnan.anwar@deakin.edu.au}

\author[1]{Haftu Tasew Reda}[orcid=0000-0001-7769-3384]
\ead{haftu.reda@deakin.edu.au}

\author[2]{Nasser Hosseinzadeh}
\ead{nasser.hosseinzadeh@deakin.edu.au}

\address[2]{Centre for Smart Power and Energy Research (CSPER), School of Engineering, Deakin University, Geelong, VIC 3216, Australia}

\author[2]{Shama Naz Islam}[orcid=0000-0002-2354-7960]
\ead{shama.i@deakin.edu.au}

\author[3]{Abdun Naser Mahmood}[orcid=0000-0001-7769-3384]
\ead{A.Mahmood@latrobe.edu.au}

\address[3]{Department of Computer Science \& IT, Latrobe University, Bundoora, VIC 3086, Australia}

\author[1]{Robin Doss}
\ead{robin.doss@deakin.edu.au}

\cormark[1]

\begin{abstract}
With the growing concern about the security and privacy of smart grid systems, cyberattacks on critical power grid components, such as state estimation, have proven to be one of the top-priority cyber-related issues and have received significant attention in recent years. However, cyberattack detection in smart grids now faces new challenges, including privacy preservation and decentralized power zones with strategic data owners. To address these technical bottlenecks, this paper proposes a novel \textbf{Fed}erated Learning-based privacy-preserving and communication-efficient attack detection framework, known as \textit{FedDiSC}, that enables \textbf{Di}scrimination between power \textbf{S}ystem disturbances and \textbf{C}yberattacks. Specifically, we first propose a Federated Learning approach to enable Supervisory Control and Data Acquisition subsystems of decentralized power grid zones to collaboratively train an attack detection model without sharing sensitive power related data. Secondly, we put forward a representation learning-based Deep Auto-Encoder network to accurately detect power system and cybersecurity anomalies. Lastly, to adapt our proposed framework to the timeliness of real-world cyberattack detection in SGs, we leverage the use of a gradient privacy-preserving quantization scheme known as DP-SIGNSGD to improve its communication efficiency. Extensive simulations of the proposed framework on publicly available  Industrial Control Systems datasets demonstrate that the proposed framework can achieve superior detection accuracy while preserving the privacy of sensitive power grid related information. Furthermore, we find that the gradient quantization scheme utilized improves communication efficiency by 40\% when compared to a traditional federated learning approach without gradient quantization which suggests suitability in a real-world scenario. 
\end{abstract}



\maketitle

\section{Introduction}

The widespread adoption of Internet of Things (IoT) in the Smart Grid (SG) paradigm has spawned an unprecedented growth in data volumes generated by edge devices. Such data are particularly vital for numerous key operations of power systems including demand forecasting, grid optimization, monitoring and control of power systems, etc. As such, the overall energy grid system management can be performed more dynamically and efficiently. However, the increase in heterogeneity, diversity, and complexity within the power system infrastructure poses a critical challenge to its integrity. Ensuring data integrity is one of the core missions of modern power systems \cite{mahusnoofdisurvey}. This is because the ubiquity of modern cyber-physical power grid components coupled with advancements in information technology now enables data processing and decision making at the edge \cite{https://doi.org/10.48550/arxiv.2203.00219} as compared to traditional cloud storage-based control centres. 

The progress of digital technology has brought about major improvements in the energy critical infrastructure, thus alleviating the modern world's most crucial problem of how to reconcile the insatiable demand for energy. Despite its significant benefits, SGs still face many technical challenges. Although smart sensing devices play a critical role in the SG paradigm, cyber attacks can impose a serious impact on the integrity and resilience of the system \cite{7045410}. Many of the SG's cyber-physical components become potential targets of  cyber attacks \cite{6900095}. One such critical cyber attack is \textit{False Data Injection} (FDI), a critical cyber attack in which malicious adversaries launch severe attacks by corrupting raw sensor measurements or state estimation outcomes \cite{Reda_Anwar_Mahmood_2022}. Furthermore, attackers constantly improve their attack crafting methods to disguise the compromised measurements as natural disturbances/faults such that the detection between malicious and non-malicious events becomes overwhelming and infeasible for bad data detection methods \cite{Anwar_Mahmood_2016}. Therefore, it is imperative to design effective countermeasures to defend modern power systems against such types of cyber threats.

In the recent past, several studies have been conducted to develop defenses against cyber attacks on SGs. Previous works have focused on leveraging machine learning algorithms for cyberattack detection in smart grid systems. For example, the work in \cite{6880823} utilizes a supervised learning-based distributed support vector machine along with alternating direction method of multipliers and achieved sufficient detection performance and lower computation complexities when tested on IEEE standard test systems. Another literature by Anwar, Mahmood and Shah \cite{10.1145/2806416.2806648} claims that their solution based on Principal Component Analysis and Sequential Minimal Optimization (SMO)-based support vectors outperformed popular machine learning algorithms when tested on industrial control systems datasets. More recently, deep learning approaches have been employed for cyber attack detection due to their superior accuracy against conventional machine learning algorithms. For instance, in \cite{8623514}, the authors developed an unsupervised deep belief network-based detection method which led to satisfactory results when tested on IEEE standard test system. On the other hand, the work in \cite{9833475} proposes a state-of-the-art deep ensembled learning algorithm to improve FDI detection in a residential demand response context. However, centralized cyber attack detection techniques proposed in numerous previous literature suffer from: 1) limited storage capabilities, 2) communication bottlenecks, and most importantly, 3) \textit{privacy} issues. Researchers often overlook the fact that power system data may contain sensitive information and its communication to centralized control centres may eventually lead to privacy leakages \cite{https://doi.org/10.48550/arxiv.2209.14547}.

To address the aforementioned challenges, a promising privacy-preserving  distributed learning-based cyber attack detection method that leverages Federated Learning (FL) was proposed. This new machine learning paradigm enables edge devices to train a cyber attack detection deep learning neural network model on-device at the edge with their local training samples, all without requiring the sharing of their local dataset without compromising privacy. For example, the work in \cite{9547719} proposed a FL framework for detecting FDI attacks on solar PV DC/DC and DC/AC converters. Similarly, \cite{linbun} developed an incentive-based federated attack detection on power grid state estimations which is claimed to outperform existing works. The authors in \cite{9878267} implemented a federated transformer-based FDI detection system followed by extensive validations on IEEE standard test systems which revealed the effectiveness of their proposed mechanism. Unfortunately, there are limited number of studies that focus on FL for detecting stealthy cyber threats on power control systems.

In this work, we propose the suitability of a federated cyber attack detection mechanism as a means to discriminate between malicious and non-malicious power system disturbances which relieves the burden of a power system operator to determine whether a certain power disturbance is intentional and thus, provide reliable decision support. We set up distributed power grid zones whereby SCADA sub-systems can locally detect cyber threats instead of uploading data to the control centre, thereby addressing storage capability, communication reliability and privacy issues faced by centralized approaches. In particular, the main contributions of this paper are summarized as follows: 
\begin{enumerate}[wide, labelwidth=!, labelindent=0pt]
\item We propose a light-weight privacy-preserving on-device collaborative deep cyber attack detection framework for power  system networks based on federated learning, known as \textit{FedDiSC}. 
\item We put forward a representation learning-based Deep Auto-Encoder (DAE) network to accurately detect power system and cybersecurity anomalies.
\item We leverage the use of DP-SIGNSGD \cite{https://doi.org/10.48550/arxiv.2105.04808}, a privacy-preserving gradient quantization scheme, to improve the computation efficiency of our proposed framework. Such a scheme decreases the communication overhead by reducing the number of parameters exchanged between the client nodes and the central orchestrator.
\item We conduct extensive experiments on publicly available Industrial Control Systems datasets to validate that our proposed framework indeed detects anomalies with high accuracy whilst keeping a low communication overhead. We evaluate our proposed framework against several FL-based deep learning models and centralized machine learning models as discussed in Section \ref{sec:FDIDetectPerformance} and find that our  attack detection framework achieves a superior accuracy of 94.8\% and a 40\% decrease in training time.
\end{enumerate}

The remainder of this paper is structured as follows: The previous state-of-the-art literature in regards to cyber attack detection and Federated Learning is presented in Section \ref{sec:relatedwork}. The proposed \textit{FedDiSC} approach, including the system model, detection module, etc. is presented in Section \ref{sec:proposed Methodology}. Simulation scenarios for validating the effectiveness of our proposed approach on publicly available datasets are presented in Section \ref{sec:results}. Finally, we conclude the paper in Section \ref{sec:conclusion}.

\section{Related Works}
\label{sec:relatedwork}

In what follows, we summarize the current state-of-the-art research on cyber attack detection and FL by classifying the previous related literature into two main categories: Data-driven cyber attack detection and Federated Learning in relation to SGs.

\subsection{Data-driven Cyber Attack Detection}

Whilst there are several approaches and directions in cyber attack detection, model-based detection algorithms have garnered huge attention in literature due to their scalability, real-time detection capabilities and non-dependence of system parameters \cite{Majidi_Hadayeghparast_Karimipour_2022}.  For instance, the work in \cite{6033036} is one of the earliest to investigate the use of a Gaussian process regression modelling technique based on prior information to develop an early warning system to mitigate the impact of cyberattacks in smart grids. Another work in \cite{6900095} trained popular machine learning models such as Random Forests, Naive Bayes, etc. on industrial control datasets and achieved high detection performance.  The same   further explored the use of different benchmark classifiers such as Random Forest, J48, JRip, etc. in \cite{8691899}. Hao, Piechocki, Kaleshi, Chin and Fan\cite{7234893} proposed the use of a Robust Principal Component Analysis (RPCA) approach for both random and targeted FDI attack detection. In similar line, Yang, Wang, Zhou, Ruan  and Liu \cite{Yang_Wang_Zhou_Ruan_Liu_2018} employed four conventional machine learning models, namely one-Class SVM, Robust covariance, Isolation forest and Local outlier factor method for cyber attack detection and achieved high classification performance after simulation on an IEEE 14-bus system. Another recent study undertaken by Kumar, Saxena and Choi \cite{9333913} addressed the problem of cyberattack detection by training several popular machine learning models such as AdaBoost, Gradient Boosting, Nearest Neighbours, etc. on a power system dataset to achieve detection accuracy ranging between 70\% and 92\%. 

In recent years, with the exponential rise of deep learning applications, researchers are actively utilizing newer deep learning algorithms to further advance and improve cyber threat detection in smart grids. For example, the work in \cite{Anwar_Mahmood_Ray_Mahmud_Tari_2020} developed a multi-layer perceptron based approach for improving the resilience of SG systems against FDI attacks. Another work by Niu, Li, Sun and Tomsovic \cite{8791598} developed a CNN-LSTM network for anomaly detection in smart grids which outperformed traditional bad data detectors. Similarly, the authors in \cite{9049087} employ a state-of-the-art CNN as a multi-label classifier to defend against potential corrupted measurement threats. Wang, Zhang, Ma and Jin \cite{9542963} proposed the use of a Kalman filter paired with a recurrent neural network demonstrated significantly improved detection rates against cyber attacks on smart grids. More recently, Bitirgen and Filik \cite{Bitirgen_Filik_2023} proposed a CNN-LSTM ensembled model combined with particle swarm optimization which achieved  high cyberattack detection accuracy rates as compared to existing works.However, there is a privacy issue when separate power grid zones are managed by various system operators, which is a very common scenario in power system networks. And, much of the above literature do not address the problem of user privacy when dealing with such sensitive power related data for detection of cyber threats. Unlike existing research works, our proposed privacy-preserving gradient quantization scheme combined with FL addresses this privacy issue by enabling a computationally-efficient on-device collaborative training of the attack detection model for discriminating between power system disturbances and cyber attacks.

\subsection{Federated Learning in smart grid applications}

FL is a fundamentally novel machine learning paradigm that enables collaborative decentralised on-device training of models without requiring the physical migration of raw data to a centralized server \cite{Fedpap}. Since its inception, FL has been applied to several SG applications where privacy is paramount. Taik and Cherkaoui \cite{9148937} pioneered the application of FL in the smart grid domain by designing a privacy-preserving demand load forecasting framework that required no data sharing. They trained a LSTM model on a real-world Texas load consumption dataset and achieved sufficient forecasting performance to claim the effectiveness of FL in the smart grid domain. Another study in \cite{9729772} leveraged FL applications for distributed energy resources forecasting and validated their high forecasting performance using a GridLAB-D simulations and Pecan Street dataset. Similarly, Gholizadeh and Musilek \cite{Gholizadeh_Musilek_2022} and Yang, Wang, Zhao and Wu \cite{Yang_Wang_Zhao_Wu_2023} achieved sufficient load prediction performance after employing FL for demand forecasting. Due to its prominence and optimal performance in the demand forecasting application, FL quickly garnered traction in other aspects of the smart grid domain. For example, the authors in \cite{9531953} designed a federated temporal convolutional network for energy theft detection. They claimed that, after extensive data-driven experiments using a real-energy consumption dataset, their federated framework can achieve high accuracy of detection with a smaller computation overhead. On similar line, Ashraf, Waqas, Abbas, Baker, Abbas and Alasmary \cite{Ashraf_Waqas_Abbas2022} achieved comparable performance in energy theft detection after employing federated conventional machine learning algorithms on real-world datasets. More recently, there has been few works in relation to federated cyberattack detection. The work in \cite{9547719} proposed a FL framework for detecting FDI attacks on solar PV DC/DC and DC/AC converters. Similarly, \cite{linbun} developed an incentive-based federated FDI attack detection on power grid state estimations which is claimed to outperform existing works. The authors in \cite{9878267} implemented a federated transformer-based FDI detection system followed by extensive validations on IEEE standard test systems which revealed the effectiveness of their proposed mechanism. In contrast, our focus in this paper constitutes of developing a federated grid-wide anomaly detection framework for power control systems that enables the discrimination between non-malicious power system disturbances and malicious FDI threats whilst being computationally efficient.

\section{Proposed Methodology}
\label{sec:proposed Methodology}
In this section, we will first present the system model considered for this scenario, followed by our proposed computationally-efficient FL-based privacy-preserving intrusion detection framework. 

\subsection{System Model and Assumptions}

\begin{figure*}
    \centering
    \includegraphics[width=16.5cm]{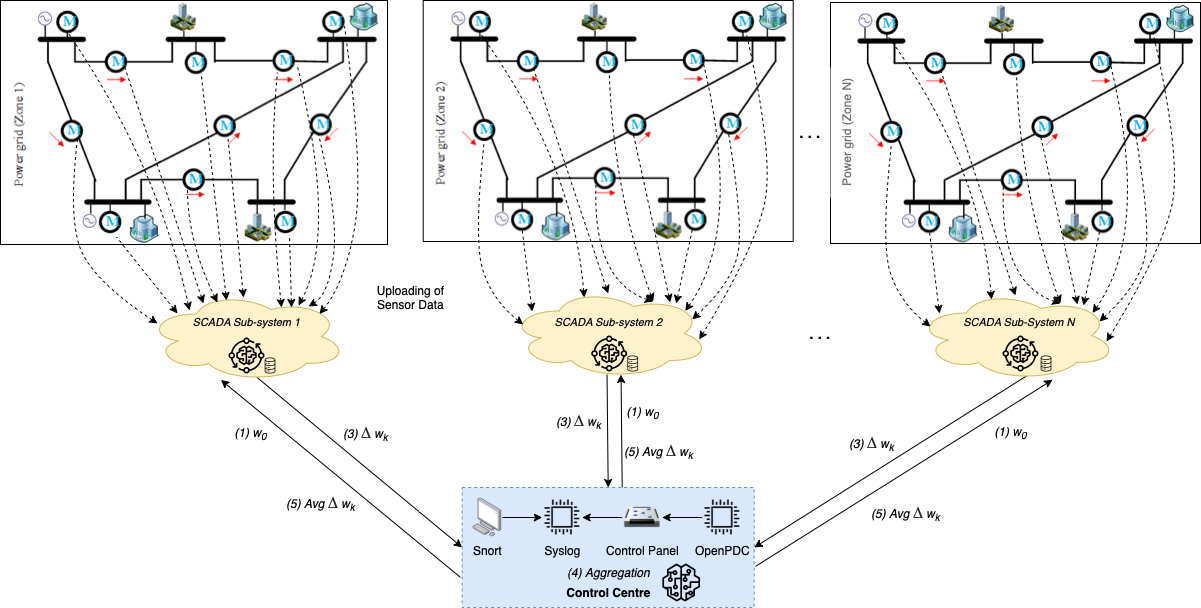}
    \caption{An overview of the FL-based intrusion detection framework. An intuitive workflow of our proposed framework is as follows: 1) The edge IEDs in power grid zones uploads its data securely to a SCADA sub-system. 2) The control centre distributes unanimous model and parameter initialization to all SCADA sub-systems. 3) Local models are trained by each SCADA sub-system with their own local dataset. 4) Each SCADA sub-system updates its compressed gradient to the control centre. 5) The control centre aggregates the uploaded compressed gradients to obtain the latest model. 6) The control centre sends back the model updates to each SCADA sub-system. The steps 1-6 are then repeated until the global model reaches optimal convergence. \label{fig:SystemFLModel}}
\end{figure*}

Throughout this manuscript, we consider a generic zone-wise distributed power grid system, which involves a control centre and $K$ edge-based Supervisory Control and Data Acquisition (SCADA) sub-systems that collaboratively train an FDI attack detection model as illustrated in Figure \ref{fig:SystemFLModel}. The edge-based SCADA sub-systems train the aforementioned shared model locally using their own local dataset, which is collected from the different IED sensing nodes in each respective power grid zones. After local training, the model updates are then shared to the control centre which uses majority-voting aggregation algorithm to average the model updates, thus obtaining an updated global model. The updated model gradients are then shared back to the edge SCADA sub-systems, which will in turn be used for FDI attack detection. The three key actors present in our proposed system model are briefly described as follows: \\ 1) \textit{Control Centre}: The control centre is the central facility that monitors and coordinates the grid operations. Typically, the control centre has cloud servers and computational resources which provide sufficient computation power to act as the central orchestrator of the FL set-up. It will be mainly responsible for initializing the global model, broadcasting the model parameters to the edge-based SCADA sub-systems and lastly, aggregating the shared model gradients from the edge-based clients. We assume that the control centre is an honest-but-curious participant, which means that the control centre attempts to learn as much a possible without acting adversarial.\\ 2) \textit{SCADA Sub-systems}: The SCADA sub-systems are edge-based clients with advanced data acquisition capabilities that play a significant role in the collection and monitoring  of data sensed by each sensor device (i.e., IED) within a power grid system. The SCADA sub-systems will act as the FL nodes whereby they will train a local model using their own local dataset and update the compressed gradients to the control centre. Furthermore, we assume that each SCADA sub-system is honest and not curious, in which case they will not be communicating with each other laterally.\\ 3) \textit{Power Grid Zones}: The power grid zones are considered to be equipped with decentralized monitoring of the SG which consist of sensor networks and IEDs (such as relays, synchrophasors, and so on). The IEDs across the physical power grids constantly sense grid-related data including voltage phase angle, current magnitude, relay status, and so on. The IEDs are assumed to be connected to the SCADA sub-system using high-speed communication network to transmit sensed data frequently. 

\subsection{Proposed Attack Detection Module}
\begin{figure*}
    \centering
    \includegraphics[width=16.5cm]{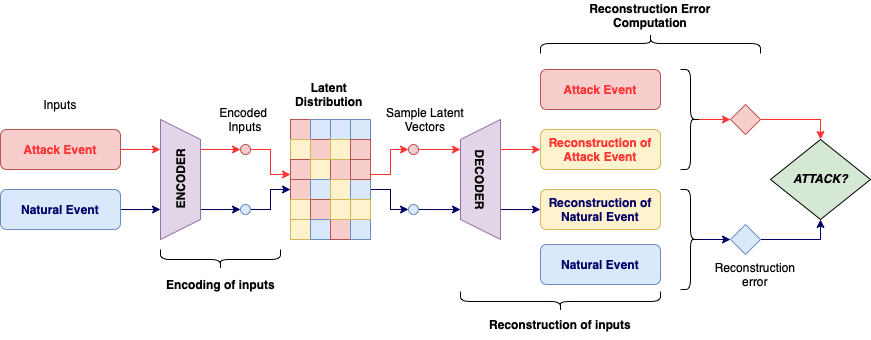}
    \caption{A schematic overview of our proposed attack detection module architecture with an example of a data sample going through the anomaly detection process. The detection module consists of three components: 1) \textit{Encoder}: The encoder compresses the input into a lower dimensional latent space representation. 2) \textit{Latent Distribution}: By imposing a bottleneck in the network, we enforce a compressed knowledge representation of the original input such that sample latent vectors can be fed to the decoder. 3) \textit{Decoder}: The decoder outputs a reconstruction of the original input which is then compared against the original input. \label{fig:autoencodermodel}}
\end{figure*}

The problem of cyber attack detection in power systems is formulated using a DAE model, as shown in Figure \ref{fig:autoencodermodel}. DAEs are powerful unsupervised representation learning-based neural networks whose goals are to achieve identity mapping between their inputs and outputs \cite{10.1145/3097983.3098052}. Auto-encoders learn important features of by sending an input vector, $x_i$, to an encoder, $f_\alpha$, which maps $x_j$ to a latent space representation (also commonly referred to as information bottleneck), $y^{(j)}$ which is essentially a compressed low-dimensional knowledge representation of the original input, such that $y^{(j)}=f_\alpha (x_j) = a(Zx_j + c)$, where $a$ is an element-wise activation function, $Z$ is the weight matrix and $c$ is the bias vector of the encoder.  The sample latent vector, $y_j$ is then passed to the decoder, $g_{\alpha'}$, which performs a reconstruction of the original input, $x_j$, through $x_j=g_{\alpha'}(y_i)=a(Z'y_j + c')$ while minimizing the reconstruction error. The parameters of the model $Z$, $Z'$, $c$ and $c'$ are to be adjusted throughout the federated training process such that $\alpha$, $\alpha' = arg \displaystyle{\min_{\alpha, \alpha'}} \ell(x_j, y_j)$ where $\ell(x_j, y_j)$ is the reconstruction error function to be minimized. 

The proposed DAE network architecture consists of two symmetrical deep belief networks, the first five shallow layers representing the encoding half of the network and the second set of five shallow layers representing the model's decoding part. The layers consist of Restricted Boltzmann Machines (RBMs), which are the building blocks of deep belief networks. Each RBM layer allows communication between its previous and subsequent layer, i.e., it serves as a hidden layer to the previous nodes and as a visible layer to the subsequent nodes. However, there is no lateral communication between nodes of a single layer. The RBMs are trained using the Contrastive Divergence Algorithm \cite{5596468}. In addition, all the DAE layers used rectified linear unit (ReLU) activation function as it improves the performance of RBMs, is easier to train and often achieves better performance. The reconstruction error is a description of the existence of cyber attacks. In our scenario, the reconstruction error metric selected is the Mean Squared Error (MSE) which acts as a measurement of anomaly degree. If the reconstruction error, $r$, exceeds a certain threshold $\tau$, then a cyber attack may have occurred. At the end of the training process, the reconstruction errors for the training set are sorted in ascending order and the value where an inflection point occurs in the error distribution is chosen as the value for $t$. For the purpose of classification, we add soft-max layers at the end of the RBM stack. The batch size and learning rate for training was set to 100 and 10$^{-3}$ respectively. This model serves as the basis for FDI attack detection in this manuscript. 
\subsection{Privacy-preserving Gradient Quantization Mechanism}
\label{GradQuan:sec}
During training, the goal of the DAE is to optimize the objective function, $g(x) = arg \displaystyle{\min_{\alpha, \alpha'}} \ell(x_j, y_j)$ such that the reconstruction error is minimized. Typically, the search for the optimal DAE parameters for any given identity-mapping task, $f:M \rightarrow N$, within an FL set-up, can be accomplished through the use of Stochastic Gradient Descent (SGD) algorithm which chooses the loss of random training samples to approximate the objective function. However, as depicted in Figure \ref{fig:SystemFLModel}, large-scale federated training for cyber attack detection in smart grid systems requires frequent bi-directional communication of local model updates between the control centre and the SCADA sub-systems. The size of the local model updates increases the communication bandwidth which limits the scalability of our proposed approach to larger number of client nodes. Furthermore, the context of local model gradients can leak privacy through easy accessibility, thus not conforming to the privacy-preserving guarantees of FL \cite{https://doi.org/10.48550/arxiv.2209.14547}.

\begin{algorithm}
\textbf{Input}: learning rate $\eta$, one-bit quantizer $sign(.)$, $K$ SCADA sub-systems with independent gradients $g^t_k$, current weight parameter $Z^{t}$.
\bigbreak
\textbf{SCADA sub-system}:

\hskip0.5em \textit{Compute} normalized gradient $\bar{g}^t_k = g^t_k - \mu^t_k$.
\vskip0.5em
\hskip0.5em \textit{Compute} quantized gradient $\hat{g}^t_k =dpsign(\bar{g}^t_k)$.
\vskip0.5em
\hskip0.5em \textit{Send} $\hat{g}^t_k$ to control centre.
\bigbreak
\textbf{Control Centre}:

\hskip0.5em \textit{Obtain} gradient $\hat{g}^t_k$ from each SCADA sub-system $k$.
\vskip0.5em
\hskip0.5em \textit{Compute} new gradient $\hat{g}_{MV}$ through majority-vote based aggregation of $\hat{g}^t_k$ for all $K$ such that $\hat{g}_{MV}$ = $sign \left(\displaystyle \sum^K_{k=1} \hat{g}^t_k \right)$.
\vskip0.5em
\hskip0.5em \textit{Update} new model parameters $Z^{t+1}$ such that $Z^{t+1}$ = $Z^{t} - \eta^t \hat{g}_{MV}$.
\vskip0.5em
\hskip0.5em \textit{Send} $Z^{t+1}$ to each $k$.
\vskip0.1em
\caption{Gradient Quantization Mechanism}
\label{algo:gradientcompression}
\end{algorithm}

Therefore, to further improve the communication efficiency of our proposed framework, we propose the use of a gradient quantization approach to reduce the communication bandwidth through the approximation of gradients to low-precision values. Specifically, we leverage the use of the DP-SIGNSGD \cite{https://doi.org/10.48550/arxiv.2105.04808} algorithm whereby after local model training, each SCADA sub-system, $k$ performs a zero-mean normalization of its local gradient, $g^t_k$, such that $\bar{g}^t_k = g^t_k - \mu^t_k$ in order to make the one-bit quantized outputs from all $K$ uniformly distributed. Each $k$ then compresses its local normalized gradient $\bar{g}^t_k$ by utilizing a one-bit differentially-private quantizer which simply takes the sign of $\bar{g}^t_k$, such that $\hat{g}^t_k = dpsign(\bar{g}^t_k)$. Each SCADA sub-systems then sends its quantized gradients, $\hat{g}^t_k$ to the control centre instead of its actual local gradient $g^t_k$. On the other hand, to attain a one-bit compressed global gradient estimate, the control centre applies a majority-vote based aggregation method as detailed in \cite{pmlr-v80-bernstein18a} to takes the sign of the summation of the one-bit local gradients such that $\hat{g}_{MV} = sign \left(\displaystyle \sum^K_{k=1} \hat{g}^t_k \right)$.  Following the estimation of $\hat{g}_{MV}$, the control centre updates the new weight parameters of the model as $Z^{t+1} = Z^{t} - \eta^t \hat{g}_{MV} $ where $\eta^t \in (0,1)$ is the learning rate of the SGD algorithm. This gradient quantization technique combined with majority-voting for model updates aggregation, as presented in Algorithm \ref{algo:gradientcompression} reduces the byte size of the gradient updates exchanged between the SCADA sub-systems and the control centre to enhance communication efficiency.

\subsection{Federated Learning Framework}

As illustrated in Figure \ref{fig:SystemFLModel}, we consider an FL scenario whereby a a control centre is connected to $K$ SCADA sub-systems via a communication systems (such as the IEC 61850 protocol). In our proposed framework, we assume that the data communication protocol between the SCADA sub-system and the control center is secure and trustworthy. The primary objective of of the FL scenario is to collaboratively train the global DAE model accross the different SCADA sub-systems using their own local dataset without having to upload sensitive power related data to the control centre. Each SCADA-subsystem dataset is acquired from multiple IEDs present within the respective power grid system zone. Let $D_k = \{a_k^i, b_k^i\}^{N_k}_{i=1}$ be the training dataset for a particular SCADA-subsystem $k \in [K]$ where $N_k = |D_k|$ is the number of data points for each $D_k$. It is important to note that the data distributions are Non-IID, as with a real-world cyber attack detection scenario. Therefore, the objective function $f_k$ for a shared global DAE model parameter $Z \in \mathbb{R}^M$ can be defined as $f_k(Z) = \dfrac{1}{N_k} \displaystyle \sum_{i=1}^{N_k}\ell \left(\{a_k^i, b_k^i\}; Z \right)$, where $\ell (.): \mathbb{R}^M \times \mathbb{R} \rightarrow \mathbb{R}$ is the loss to be optimized for the data point $\left(a_k^i, b_k^i\right)$. Furthermore, the global objective function $F$ over all the distributed $K$ datasets is defined as $F(Z) = \displaystyle \sum_{k=1}^K \dfrac{N_k}{N}f_k(Z)$ which is the sum of the local loss functions for each $k$, where $N$ is the total data points for training.

\begin{algorithm}
\textbf{Input}: learning rate $\eta$, one-bit quantizer $sign(.)$, $K$ SCADA sub-systems, Local training set $D_k$ for $k \in [1, K]$
\vskip0.5em
\hskip0.5em \textit{Initialize} common unanimous model parameter $Z_0$.
\bigbreak

\hskip0.5em \For{each communication round $t$ in $T_{cl} \in (1,n)$}{
\bigbreak
\textbf{Control Centre}:

\hskip0.5em \textit{Obtain} gradient $\hat{g}^t_k$ from each SCADA sub- system $k$.
\vskip0.5em
\hskip0.5em \textit{Compute} new gradient $\hat{g}_{MV}$ through majority-vote based aggregation of $\hat{g}^t_k$ for all $K$ such that $\hat{g}_{MV}$ = $sign \left(\displaystyle \sum^K_{k=1} \hat{g}^t_k \right)$.
\vskip0.5em
\hskip0.5em \textit{Update} new model parameters $Z^{t+1}$ such that $Z^{t+1}$ = $Z^{t} - \eta^t \hat{g}_{MV}$.
\vskip0.5em
\hskip0.5em \textit{Send} $Z^{t+1}$ to each $k$.
\vskip0.1em

\bigbreak

\textbf{SCADA Sub-System}:

\hskip0.5em \For{each $k \in [K]$}{

\textit{Compute} local gradient $g^t_k = \nabla f_k(Z^t)$ using $D_k$. 
\vskip0.5em
\textit{Compute} normalized gradient $\bar{g}^t_k = g^t_k - \mu^t_k$.
\vskip0.5em
\textit{Compute} quantized gradient $\hat{g}^t_k = dpsign$ $(\bar{g}^t_k)$.
\vskip0.5em
\textit{Update} $\hat{g}^t_k$ to control centre. 
\vskip0.5em
\textbf{end}
}

}
\textbf{end}
\caption{Proposed FL Framework}
\label{algo:federated}
\end{algorithm}

During a specific $t^{th}$ communication round, our proposed FL framework detailed in Algorithm \ref{algo:federated} consists of five specific implementation phases as in the following: 1) \textbf{Downlink Communication}: The control centre distributes the global DAE model parameter $Z^t$ to all $K$ SCADA-subsystems. For the sake of this work, we assume a secure latency and error-free communication whereby all $K$ SCADA-subsystems receive the correct initialized model parameter $Z^t$ at similar time. 2) \textbf{Local Model Training}: Each SCADA-subsystem $k$ for $k \in [K]$ computes a local gradient $g^t_k = \nabla f_k(Z^t)$ using its local training dataset $D_k \in [K]$ and $Z^t$. 3) \textbf{Gradient Quantization}: Each $k \in [K]$ performs a zero-mean normalization of its local gradient $g^t_k$ such that $\bar{g}^t_k = g^t_k - \mu^t_k$. The normalized gradient $\bar{g}^t_k$ is then quantized using DP-SIGNSGD as detailed in Section \ref{GradQuan:sec} and Algorithm \ref{algo:gradientcompression} such that $\hat{g}^t_k = dpsign(\bar{g}^t_k)$. 4) \textbf{Uplink Communication}: Each SCADA-subsystem $k$ for $k \in [K]$ updates its locally-computed one-bit quantized gradient to the control centre over the wireless communication protocol. 5) \textbf{Model Aggregation}: Once the control centre receives $\hat{g}^t_k $ from each $k \in [K]$, it performs a majority-vote based aggregation and summation of all the one-bit $\hat{g}^t_k $ to obtain the global gradient estimate $\hat{g}_{MV}$. The control centre then updates the new weight parameter $Z^{t+1}$ to each SCADA-subsystem $k \in [K]$. 

\section{Experimental Results and Discussions}
\label{sec:results}

The cyber attack detection performance of each individual SCADA sub-system is directly correlated to the overall performance of our proposed framework, as discussed in Section \ref{sec:proposed Methodology}. Thus, in this section, we evaluate the performance of our proposed architecture in terms of cyber attack detection performance, and communication efficiency.

\subsection{Dataset and Feature Space Preparation}
\label{datasetsec}
We evaluate our proposed architecture using the Mississippi State University and Oak Ridge National Laboratory Power System Attack (MSU-ORNL PS) Dataset \cite{Tommy_Morris}. It consists of 128 features and 1 target marker variable with data logs collected from four Phasor Measurement Units (PMUs). A Real-Time Digital Simulator (RTDS) was used to simulate the power system scenario implementation and data generation within a hardware-in-the-loop test bed configuration. Each PMU collected real-time synchrophasor measurement data which includes frequency, current phase angle, current phase magnitude, voltage phase angles, voltage phase magnitude, etc. at a rate of 120 samples per second. For this research, based on the dataset, we assume that an adversary is able to penetrate the power grid zones and launch cyber attacks from within the operational network of each power grid zones. The dataset consists of 15 data files which are binary classes whereby scenarios of Astem disturbances (marked as \textit{Natural Events}) and power systems cyber attacks (marked as \textit{Attack Events}) have been simulated. The cyberattacks include FDI attack and tripping command injection attack, as used in \cite{Tommy_Morris}.  

To deal with missing values present within the dataset (which may be a result of SCADA communication failure \cite{Anwar_Mahmood_Pickering_2016}), we use a traditional machine learning-based K-Nearest Neighbour (KNN) missing value imputation to avoid information loss instead of downsampling. Detection performance typically depends on the appropriate selection of the basis of the feature space. Therefore, we apply Principal Component Analysis (PCA) algorithm to cull out some features of the data that are difficult to reconstruct such that the remaining features can be represented through a compressed low-dimensional knowledge representation with minimal reconstruction error. A basis of 100 features is determined. Next, we normalize the input data into the range of (0,1) to accelerate model convergence. Lastly, we sample 30\% of the dataset for testing the accuracy of the global FDI detection model while the remaining training data is divided amongst distributed power grid zones for FL local model training.  

\subsection{Attack Detection Performance}
\label{sec:FDIDetectPerformance}

\begin{figure}
    \centering
    \includegraphics[width=8cm]{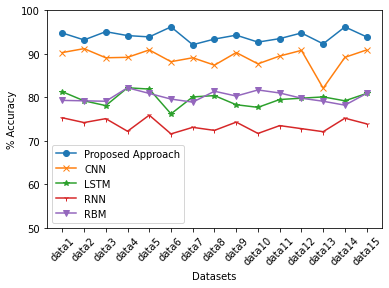}
    \caption{Comparison of the average performance of five deep learning attack detection models over the fifteen data files of the dataset. The results demonstrate the superiority of our proposed framework in attack detection over fifteen data files. \label{fig:compamodelperformaccdata}}
\end{figure}
\begin{figure}
    \centering
    \includegraphics[width=8cm]{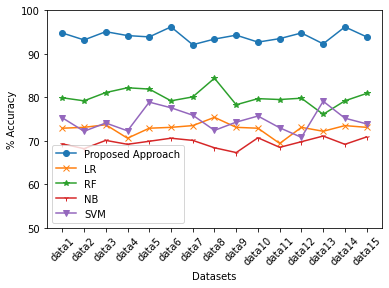}
    \caption{Comparison of the average performance of proposed DAE model vs other traditional machine learning models over the fifteen data files of the dataset. Similarly, the results demonstrate the superiority of our proposed framework in attack detection over fifteen data files.\label{fig:compamodelperformaceMLDAE}}
\end{figure}
\begin{figure}
    \centering
    \includegraphics[width=8cm]{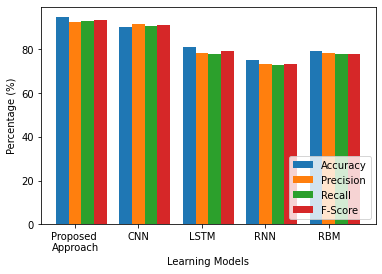}
    \caption{Comparison of the average performance of four other deep learning attack detection models against our proposed approach over the whole dataset based on the percentage accuracy, precision, recall and F-Score metrics. The results demonstrate the superiority of our proposed framework in attack detection over the whole dataset. \label{fig:metriccompa}}
\end{figure}
\begin{figure}
    \centering
    \includegraphics[width=8cm]{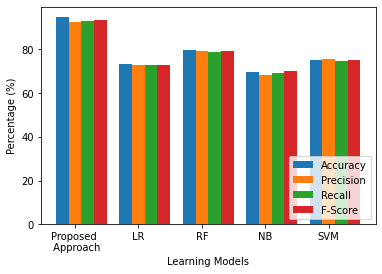}
    \caption{Comparison of the average performance of four traditional machine learning attack detection models against our proposed approach over the whole dataset based on accuracy, precision, recall and F-Score metrics. The results demonstrate the superiority of our proposed framework in attack detection over the whole dataset. \label{fig:metriccompaML}}
\end{figure}
\begin{table*}
\begin{tabular}{|l|c|c|c|c|c|}
\hline
\diagbox{\textbf{Metric}}{\textbf{Learning Model}} & \textbf{Proposed Approach} & \textbf{CNN} & \textbf{LSTM} & \textbf{RNN} & \textbf{RBM} \\ \hline
\textbf{Accuracy}  (\%)                                                                 & 94.8                     & 90.3                     & 81.3                      & 75.3                     & 79.3                     \\ \hline
\textbf{Precision} (\%)                                                                 & 92.7                     & 91.8                     & 78.5                      & 73.5                     & 78.2                     \\ \hline
\textbf{Recall} (\%)                                                                    & 92.8                     & 90.9                     & 78.1                      & 73.0                     & 77.7                     \\ \hline
\textbf{F-Score} (\%)                                                                    & 93.6                     & 91.1                     & 79.2                      & 73.2                     & 77.9                     \\ \hline
\end{tabular}
\caption{Comparison of the average attack detection performance of various deep learning  models to our proposed DAE over the whole dataset. We compare our proposed DAE model to CNN, LSTM, RNN and RBM models, all of which are trained in a FL setup. It was found that our proposed approach outperforms all competing models in terms of attack detection performance. \label{table:metricperformcompa}}
\end{table*}
\begin{table*}
\begin{tabular}{|l|c|c|c|c|c|}
\hline
\diagbox{\textbf{Metric}}{\textbf{Learning Model}} & \textbf{Proposed Approach} & \textbf{LR} & \textbf{RF} & \textbf{NB} & \textbf{SVM} \\ \hline
\textbf{Accuracy}  (\%)               & 94.8                     & 73.3                    & 79.9                    & 69.7                    & 75.3                     \\ \hline
\textbf{Precision} (\%)               & 92.7                     & 73.1                    & 79.3                    & 68.1                    & 75.4                     \\ \hline
\textbf{Recall} (\%)                  & 92.8                     & 72.7                    & 78.9                    & 69.2                    & 74.9                     \\ \hline
\textbf{F-Score} (\%)                 & 93.6                     & 72.9                    & 79.2                    & 70.0                    & 75.1                     \\ \hline
\end{tabular}
\caption{Comparison of the average attack detection performance of different traditional machine learning  models to our proposed DAE over the whole dataset. We compare our proposed FL framework to centralized LR, RF, NB and SVM traditional machine learning models. The results demonstrate the superiority of our proposed framework in attack detection. \label{table:metricperformcompaML}}
\end{table*}

We evaluated the attack detection performance of our proposed FedDiSC framework to four other popular models used in SG cyber attack detection namely Convolutional Neural Network (CNN), Long Short-Term Memory (LSTM), Recurrent Neural Network (RNN) and Restricted Blotzmann Machine (RBM). All the competing models are trained in similar FL-based set-up and configuration. We evaluate these models on the MSU-ORNL dataset described in Section \ref{datasetsec}. Initially, we train the FL models on the training sets and we evaluate and compare the accuracy of the global models after training on the MSU-ORNL PS test sets. In Figure \ref{fig:compamodelperformaccdata}, the experimental validations reveal that the proposed DAE model achieves superior accuracy over all 15 data files. For instance, the accuracy of our proposed DAE model on data1 is 94.8\% as compared to the CNN model which is 89.9\% or the LSTM model which is 81.5\%. Similarly, for data15, the DAE model outperforms the RBM model by 13.6\% and the RNN model by 19.9\%. This shows that our proposed DAE model has improved robustness to different power grid related datasets which is due to the fact that FL enables learning of features from multiple SCADA subsystems, thereby resulting in improved robustness of the model. Furthermore, we compare the average attack performance of the global DAE model over the test sets as presented in Table \ref{table:metricperformcompa} and Figure \ref{fig:metriccompa} using four classification metrics namely accuracy, precision, recall and F-Score. From the experimental results, we observe that the our proposed model outperforms the other competing ones in all three metrics. For example, the precision of the DAE model is superior to that of the competing CNN, LSTM, RNN and RBM by 0.9\%, 14.2\%, 19.2\% and 14.5\% respectively. Likewise, the recall of our proposed model outshines those of the rivalling models by 1.9\%, 14.7\%, 19.8\% and 15.1\% respectively while the F-Score of our proposed model is superior to that of the aforementioned ones by 2.5\%, 14.4\%, 20.4\% and 15.7\% respectively. 

Furthermore, we evaluated the attack detection performance of our proposed approach to that of four popular traditional machine learning-based models that have been used in previous literature namely, Logistic Regression (LR), Random Forest (RF), Naive Bayes (NB) and Support Vector Machine (SVM). Among those competing models, only our proposed model is trained in a federated learning set-up while the remainder are centralized ones. In Figure \ref{fig:compamodelperformaceMLDAE}, we evaluated the accuracy of our proposed global DAE model against four different centralized traditional machine learning models over all 15 data files. For example, the accuracy of our proposed model on data1 is 94.8\% as compared to the LR model which is 72.9\% or the NB model which is 69.4\%. Similarly, for data 15, the accuracy of our proposed model outshines that of RF and SVM by 12.9\% and 10.0\% respectively. The experimental results, as depicted in Table \ref{table:metricperformcompaML} and Figure \ref{fig:metriccompaML}, we observe that our proposed approach results in the highest attack detection performance as compared to the other rivalling traditional machine learning models. For instance, our proposed FL approach achieves higher precision compared to LR, RF, NB and SVM by 19.6\%, 13.4\%, 24.6\% and 17.3\% respectively. This may be due to the fact that deep learning algorithms are superior to the performance of traditional machine learning models as they try to learn high-level features from data in an incremental manner and improve with larger datasets. On the other hand, our experimental validations reveal that Naive Bayes has the worst overall performance amongst all the models used for comparison purposes and achieved an attack detection accuracy of merely 69.7\%. 

Lastly, we compare the results of our proposed FL-based cyber attack detection framework against that of state-of-the-art literature within this topic. The authors in \cite{7063234} developed a hybrid intrusion detection system using common paths mining algorithm and evaluated their proposed approach on similar dataset. They achieved an overall average accuracy of 73.43\% while our proposed framework had an accuracy level of 94.8\%. Similarly, in comparison with \cite{7081776}, we achieved improved accuracy of cyber-attack detection. Therefore, we can conclude that our proposed FL framework for cyber attack and power system disturbance discrimination achieves sufficient detection accuracy as compared to other deep learning models whilst guaranteeing privacy of sensitive power grid related information and being computationally efficient.

\subsection{Computational Efficiency of Our Proposed Approach}

Subsequently, we evaluate the computation efficiency achieved by the gradient compression approach utilized in our proposed FDI attack detection framework as detailed in Section \ref{GradQuan:sec}. Specifically, we contrast the training time of the models, between the proposed FL framework with Gradient Quantization (GQ) and the traditional FL approach without GQ. We use similar models with similar hyper-parameters for comparison within both setups such that the training time of the models can be used to estimate the computational efficiency of our proposed approach. In Figure \ref{fig:commeffresults}, we provide a comparison of the computational efficiency between the proposed FL framework with GQ scheme and traditional FL approach without the GQ scheme using different models. As depicted in Figure \ref{fig:commeffresults}, we note that for all the models compared, the training time of the models decrease whenever our proposed FL framework with GQ scheme is utilized. Specifically, it can be highlighted that the training time for our GQ-based FL approach is around 40\% less than the traditional FL framework without GQ. This is due to the fact that the privacy-preserving GQ scheme utilized in our proposed FL framework decreases the byte size of the gradients shared during the control centre and SCADA subsystem to one-bit without compromising the accuracy of detection of the model as discussed in Section \ref{sec:FDIDetectPerformance}. 

\subsection{Summary of Findings}

\begin{figure}
    \centering
    \includegraphics[width=8cm]{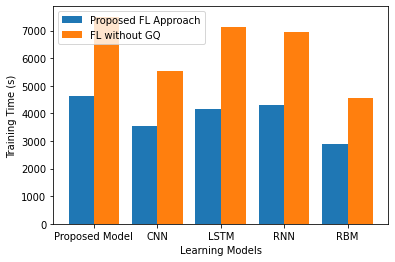}
    \caption{Training time comparison between proposed FL framework with gradient quantization scheme and traditional FL approach without gradient quantization scheme. The results of our proposed framework demonstrate a 40\% decrease in training time in comparison against other approaches. \label{fig:commeffresults}}
\end{figure}

In this study, we have focused on the detection of FDI attacks in SGs with the assumption that the SG is partitioned into $N$ number of zones where  $N \in \mathbb{R}^+$. Specifically, we proposed an FL-based cyber attack detection approach using a deep auto-encoder model for power systems disturbance and cyber attack discrimination. We evaluated our proposed approach using the MSU-ORNL PS dataset. As presented in Section \ref{sec:FDIDetectPerformance}, we observe that our proposed FL-based DAE approach achieves the best overall detection performance on the dataset as compared to other  models trained in similar fashion. In addition, we evaluated our work against centralized machine learning models which reveals that our proposed framework is superior in discriminating cyber attacks against power system faults. Similarly, a brief comparison against state-of-the-art reveals comparable performance which validates the real-world application of our proposed method for cyber attack detection. Furthermore, we improve the communication efficiency of the proposed FL architecture with a differentially private SIGNSGD-based gradient quantization scheme that conforms to the privacy-preserving nature of FL. The results, as illustrated in Figure \ref{fig:commeffresults} highlight a 40\% decrease in training time of the proposed framework as compared to a traditional FL detection model. Therefore, we can conclude that our proposed framework provides sufficient detection accuracy while improving the overall communication efficiency for FDI attack detection. 

While sufficient model performance is vital metric for effective FDI attack detection and has been achieved in our work, it is worth discussing that our proposed approach significantly differs from previous works as we also take into account the security and privacy of sensitive power related information. We restrict the training of models to only edge-based SCADA sub-systems on their local datasets which eliminates the need of data sharing to the control centre and raw data access by unauthorized attackers. Furthermore, the use of a privacy-preserving gradient quantization scheme complemented with differential privacy tackles the severe deep gradient leakage drawback suffered by FL algorithms. Adversaries are unable to get access to the raw gradients of the models shared during training and are thus, unable to extract sensitive power related information or even reconstruct the raw training data \cite{https://doi.org/10.48550/arxiv.2209.14547}. Therefore, we believe that our proposed privacy-preserving  communication-efficient framework is practical for real-world SG attack detection applications. 

\section{Conclusion}
\label{sec:conclusion}

Throughout this manuscript, we propose a novel communication efficient and privacy-preserving on-device federated learning-based intrusion detection approach for power systems disturbance and cyberattack discrimination in smart grids. First, we propose a federated learning-based framework that enables power sensing devices to collaboratively train the intrusion detection model without requiring the need for data sharing which tackles the problem of data privacy. Second, we propose a representation learning-based auto-encoder for accurate discrimination between natural power disturbances and cyberattack events. We extensively evaluate the performance of our proposed model on a popular binary power system dataset and compare it to several other machine learning/deep learning algorithms. The experimental results show that the auto-encoder model achieves superior accuracy over other methods on this particular dataset. Lastly, we propose the use of a sign-based gradient quantization scheme to enhance the communication efficiency of our proposed federated learning framework. The experimental findings show  that this gradient quantization mechanism decreases the training time by nearly 40\%, proving that quantized gradients improve communication efficiency.

In future work, we aim to improve the robustness of the power system intrusion detection models by leveraging ensembled learning methods. Furthermore, we will focus on researching techniques to improve the security of our federated learning framework, as one of our previous works \cite{https://doi.org/10.48550/arxiv.2209.14547} highlights that federated learning frameworks are vulnerable to byzantine attacks.

\bibliographystyle{unsrt}

\bibliography{cas-refs}
\end{document}